\documentclass[12pt]{iopart}

\usepackage{graphicx}
\usepackage[colorlinks]{hyperref}

\hypersetup{colorlinks,linkcolor=blue,citecolor=blue,urlcolor=blue,filecolor=blue,pdftex}

\begin{document}

\title[Stochastic coherence and synchronizability in weighted scale-free networks]{Collective stochastic coherence and synchronizability in weighted scale-free networks}

\author{Pablo Balenzuela$^{1,\dagger}$, Pau Ru\'e$^{2,3}$, Stefano Boccaletti$^4$ and Jordi Garcia-Ojalvo$^{2,3*}$}

\address{$^1$Departamento de F\'isica, Facultad de Ciencias Exactas y Naturales,
Universidad de Buenos Aires and IFIBA, CONICET , Pabell\'on 1, Ciudad Universitaria (1428), Buenos Aires,
Argentina Buenos\\
$^2$Departament of Experimental and Health Sciences, Universitat Pompeu Fabra,
Barcelona Biomedical Research Park, Dr. Aiguader 88, Barcelona, Spain\\
$^3$Departament de F\'isica i Enginyeria Nuclear, Universitat Polit\`ecnica de Catalunya, Ed. Gaia, Rbla. Sant
Nebridi 22, 08222 Terrassa, Spain\\
$^4$CNR-Istituto dei Sistemi Complessi, Via Madonna del
Piano 10, 50019 Sesto Fiorentino, Italy
}
\ead{$^{\dagger}$\mailto{balen@df.uba.ar}$, ^{*}$\mailto{jordi.g.ojalvo@upf.edu}}

\begin{abstract}
Coupling frequently enhances noise-induced coherence and synchronization in interacting nonlinear
systems, but it does so separately.
In principle collective stochastic coherence and synchronizability are incompatible phenomena,
since strongly synchronized elements behave identically and thus their response to noise
is indistinguishable to that of a single element. Therefore one can expect systems that
synchronize well to have a poor collective response to noise.
Here we show that, in spite of this apparent conflict, a certain coupling architecture is able to
reconcile the two properties. Specifically, our results reveal that weighted scale-free
networks of diffusively coupled excitable elements
exhibit both high synchronizability
of their subthreshold dynamics and a good collective response to noise of their
pulsed dynamics. This is established by comparing the behavior of this
system to that of random, regular, and unweighted scale-free networks.
We attribute the optimal response of weighted scale-free networks to the link homogeneity (with respect to node degree) provided by the weighting procedure, which balances the degree heterogeneity typical of SFNs.
\end{abstract}


\maketitle


\section{Introduction}
\label{intro}

In the face of the unavoidable randomness of nature, an
appealing hypothesis is that natural systems are optimized to use noise \cite{saguesRMP}.
A particular example of this ability is {\em stochastic coherence}, also known as coherence resonance,
a phenomenon through which noise
extracts an intrinsic time scale out of a nonlinear stochastic
system, leading to an optimally periodic (coherent) behavior for an intermediate noise level \cite{pikovsky,lindner2000coherence,revexc}.
An intuitive understanding of this effect comes from
considering a single excitable element subject to noise. Noise excites
 large-amplitude excursions (such as spikes, or action potentials, in the case
of neurons) away from, and back towards,
an otherwise stable fixed point of the system. These excursions
become more frequent for
increasing strength of the random perturbations, with the time interval
between excursions being bounded from below by a refractory
time. At an intermediate noise level spikes pile up and end up occurring almost periodically,
at intervals close to the refractory time. For larger noise levels
disorder kicks in, degrading that optimally coherent response.

Such a seemingly counter-intuitive constructive role of noise
can be further enhanced by coupling in arrays of dynamical
elements \cite{changsong}. Coupling between excitable elements
enhances stochastic coherence by ``reminding'' a given element in the array
to fire when a complying neighbor fires at the ``correct'' time (i.e.
right after the refractory period has ended). In that way,
coherence resonance is enhanced for an intermediate coupling
level: when coupling is too small, reminders do not reach the neighboring
cells; when it is too large, the array operates almost synchronously,
like a single element, and the enhancement effect naturally disappears.
The latter effect implies that one can expect strong
synchronization to be detrimental to array-enhanced stochastic coherence
\cite{nuestro}.

In the light of the preceding discussion, it would be natural to
expect that stochastic coherence is not favored in
networks with small-world properties (short path length and high clustering), since such networks
seem to favor synchronization \cite{WS}. However, it has been observed that
the intrinsic heterogeneity of small-world networks,
in which different nodes have in general different number of links,
leads in fact to a decrement in synchronizability \cite{Nishi}, in what has come
to be known as the {\em paradox of heterogeneity}.
Accordingly, stochastic coherence has been shown to persist
in small-world networks \cite{kwon}. Poor synchronizability also
occurs in standard scale-free networks, in which the
distribution of links reaching a node
(its degree) follows a power law, thus leading to substantial heterogeneity
among the nodes \cite{albert}. This limited capacity for synchronization is concurrent
with the existence of multiple instances of noise-induced coherence in these
networks \cite{acebron,Perc:2008uq,Perc:2009fk}. Thus synchronization and collective stochastic
coherence seem to be incompatible phenomena.

Here we study whether, in spite of the above-mentioned expectations,
there are network architectures
that exhibit {\em both} strong synchronizability and high levels of
stochastic coherence simultaneously. We concentrate on weighted scale-free networks,
in which the strength of the links is scaled according to the local connectivity. These
networks have been
shown to exhibit large synchronizability \cite{Bocca1,Bocca2,Motter},
but is stochastic coherence accordingly reduced in them?
Our results indicate that this is not the case, and that this coupling architecture, while still
supporting a high level of synchronizability, maintains
its ability to enhance stochastic coherence through coupling. Thus we suggest
that these weighted scale-free topologies are optimal to operate in a stochastic
environment when synchronizability is also required.

\section{Model}

We use a configuration of $N$ excitable elements (which could represent,
for instance, neurons) whose dynamics
is assumed to be given, without loss of generality,
by the FitzHugh-Nagumo model \cite{pikovsky},
\begin{eqnarray}
\epsilon \frac{dx_i}{dt} & = & x_i -\frac{x_i^3}{3} - y_i + I_i \label{eqx} \\
\frac{dy_i}{dt} & = &x_i + a + D\xi_i(t)\,, \label{eqy}
\end{eqnarray}
where $x_i$ is an activator variable and $y_i$ an inhibitor variable,
$i=1\ldots N$ labels the neurons, $a$ is a control parameter,
$\epsilon\ll 1$ is the ratio of time scales
of the activator and inhibitor, and $I_i$ is a coupling term.
The last term in Eq.~(\ref{eqy}) corresponds to a white noise
of zero mean and amplitude $D$, uncorrelated between different elements,
$\langle\xi_i(t)\xi_j(t')\rangle=2\delta_{ij}\delta(t-t')$.
In the absence of noise and coupling, the model given by
Eqs.~(\ref{eqx})-(\ref{eqy}) shows a bifurcation to a
limit cycle for decreasing $a$, at $|a|=1$. For $|a|$ slightly larger than 1, the system is excitable.
The specific values of the parameters used below are $a=1.05$
and $\epsilon=0.01$. The equations were integrated using the Heun method \cite{nises}, which corresponds to a second-order Runge-Kutta algorithm for stochastic equations.

We couple the excitable elements diffusively, representing for instance
electrical connections arising at gap junctions between pairs of neurons:
\begin{equation}
I_i = g \sum_{j=1}^Nn_{ij} (x_j-x_i). \label{elecgap}
\end{equation}
Here $g$ is the coupling strength and $n_{ij}$ are the elements
of the network connectivity matrix: $n_{ij}=0$ if $i$ and $j$ are not
connected, and {\bf $n_{ij}>0$} if they are connected.

We consider two main types of network topologies in what follows: random 
Erd\"os-Renyi (ER) networks, in which the connections are selected at random between pairs
of nodes, and scale-free networks (SFNs), in which the nodes are connected in
such a way that the distribution of degrees (number of connections that a node
has with others) follows a power law. This power-law behavior leads to a strong
degree heterogeneity among the network elements, which as mentioned above
curtails the emergence of synchronization in these networks \cite{Nishi}. The
dynamical effects of this structural heterogeneity can be balanced by weighting
the coupling strength between each pair or nodes ($i$,$j$) depending on the load
${\ell}_{ij}$ of the link connecting them. The load of a link
quantifies the traffic of shortest paths that are making use of
that link \cite{goh01}, and therefore reflects the network
structure at a global scale (its value can be strongly influenced
by pairs of nodes that may be very far away from nodes $i$
and $j$). In order to determine the loads of all links in the
network, we follow the approach of Ref.
\cite{Bocca2} and count, for each pair of nodes $(i',j')$,
the number $n(i',j')$ of shortest paths
connecting them. For each one of such shortest paths, we then add
$1/n$ to the load ${\ell}_{ij}$ of each link forming it. The
elements of the connectivity matrix for the weighted SFNs
are then given by:
\begin{equation}\label{newCouplingScheme}
   n_{ij}=  \frac{{\ell}_{ij}}{{\sum_{k \in K_i}
    {\ell}_{ik}}} ,
\end{equation}
where  $K_i$ is the set of neighbors of the $i^{\rm th}$ node
(note that this leads to an asymmetric coupling between any pair of nodes
$i$ and $j$). In ER and unweighted SFNs, in contrast, $n_{ij}=1$
for all connected node pairs. In order to do a proper comparison between networks,
we rescale the connectivity matrix $n_{ij}$ in weighted networks in such a
way that $\sum_{i,j}n_{ij}=2M$, where $M$ is the total amount of edges of the network,
as expected in unweighted networks \cite{Bocca2}.

\section{Synchronizability}
\label{sync}

\begin{figure}[htb]
\raggedleft
\includegraphics[height=0.35\textwidth,keepaspectratio,clip]{figs/subthreshold}~~~
\includegraphics[height=0.35\textwidth,keepaspectratio,clip]{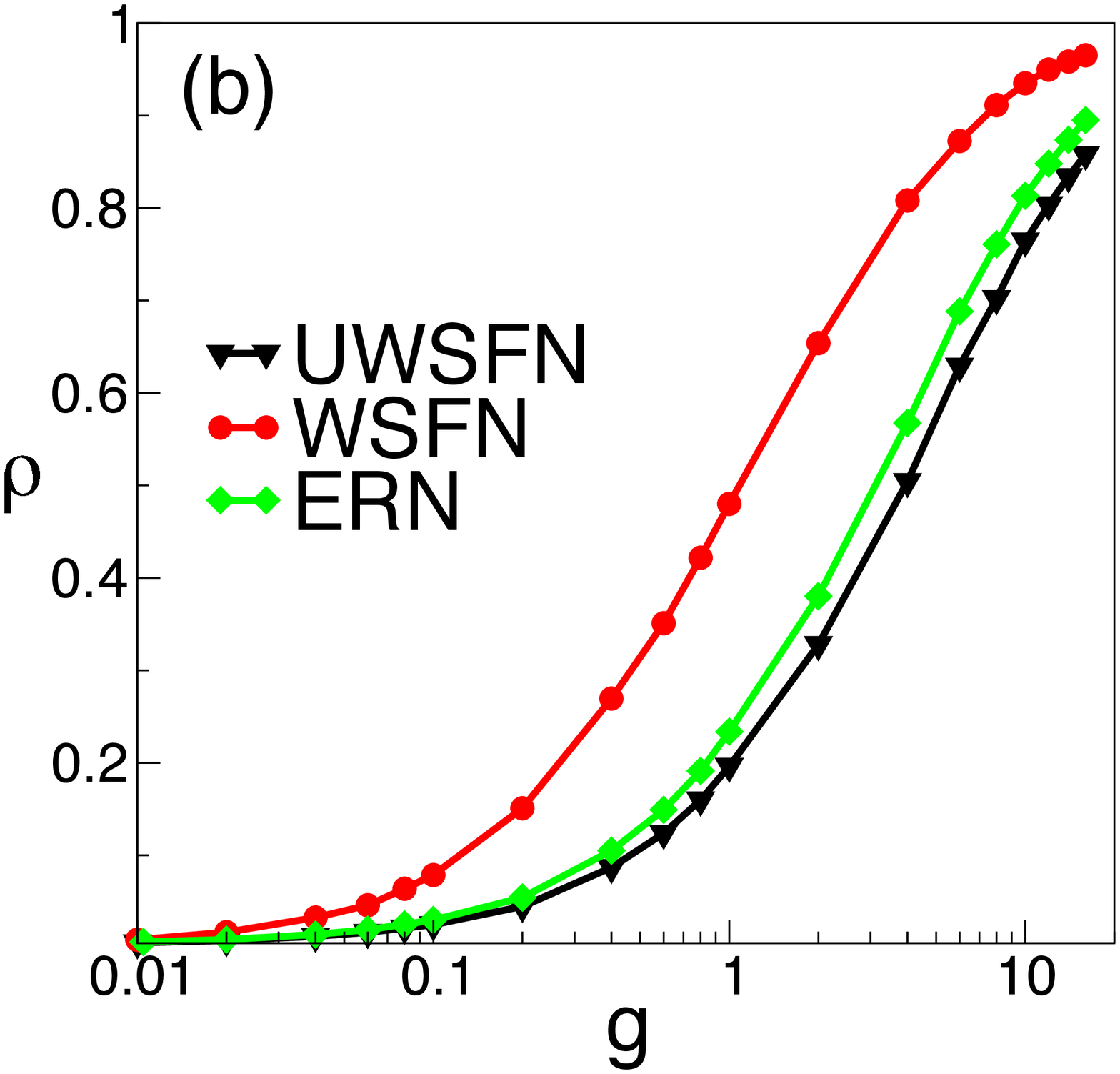}
\caption{Synchronizability of the subthreshold dynamics of
unweighted scale-free networks (black, top plot in panel (a) and triangles in
panel (b), weighted scale-free networks (red, middle plot in panel (a) and circles in
panel (b), and random networks (green, bottom plot in panel (a) and diamonds in
panel (a). Panel a shows time traces of selected individual network elements (light-shade lines) 
and of the average activity of the network (dark-shade lines). Panel (b) shows
the synchronization coefficient $\rho$ as a function of the coupling strength $g$.
The parameter values are $D=0.01$ and $g=1$. 
Curves in (b) are averages over ten network replicas of each type.
}
\label{fig:1}
\end{figure}

We first examine the subthreshold dynamics of the excitable elements described by
Eqs.~(\ref{eqx})-(\ref{eqy}). Figure~\ref{fig:1}(a)
shows (in grey lines) the temporal behavior of 11 (out of a total of $N=500$) network elements
in the absence of spiking activity,
for the three different coupling architectures described above:
unweighted (UWSFN, top) and weighted (WSFN, middle) SFNs, and random networks (ERN, bottom).
The noise intensity (the same for all three network types) is chosen low enough so that spikes are effectively absent.
In each case, the average activity of the complete network is shown superimposed to the
individual time traces, in thick lines. The amplitude of the fluctuations of that average
activity reflects the level of synchronization of the network: a large level of synchronization
between the network elements leads to an average activity that resembles that of every
single oscillator, which fluctuates due to the added noise. In the absence of synchronization,
on the other hand, the dynamics of the different oscillators average out and the fluctuations
of the average signal are reduced. Figure~\ref{fig:1}(a) shows that the average dynamics
of the weighted SFNs (middle plot) fluctuates more strongly than those of the
unweighted SFNs (top) and random networks (bottom), thus suggesting that
synchronization of the subthreshold dynamics is stronger in the latter type of
network architecture, in accordance with the synchronization properties of that
type of coupling topology discussed in Sec.~\ref{intro} above.

In order to quantify in a systematic way the synchronization capabilities of the
three types of networks, we computed a synchronization coefficient as defined in \cite{JordiPNAS2004}:
\begin{equation}\label{eq:synchro}
\rho=\frac{\langle\overline{x_i}^2\rangle-\langle\overline{x_i}\rangle^2}{\overline{\langle x_i^2\rangle-\langle x_i\rangle^2}}, 
\end{equation}
where the overlines indicate average over nodes, whereas the angle brackets $\langle ...\rangle$ indicate
temporal averages. This coefficient could be read as a the ratio between fluctuations of the global averaged
signal and the average of fluctuations of individual network elements. If the system is not synchronized, the
individual signals $x_i(t)$ will be completely out of step with respect each other and their sum will be
averaged out to zero. In the synchronized case, the fluctuations of the global signal are similar to the
fluctuations of individual neurons and the coefficient $\rho$ tends to one.

This quantifier is plotted in Fig.~\ref{fig:1}(b) as a function of $g$, showing that
all three network types exhibit a smooth transition to synchronization as coupling
increases, but the weighted SFN exhibits a larger synchronization coefficient
for all coupling levels, and thus reaches synchronization earlier as coupling
increases. A similar enhancement of synchronization is observed in the spiking
regime, provided only the subthreshold dynamics is considered (results not shown).
Therefore, weighting the connections in an SFN according to
expression (\ref{newCouplingScheme}) does lead to a higher synchronizability
than standard unweighted SFNs, and even random networks,
in spite of the structural degree heterogeneity of the network.

\section{Stochastic coherence}
\label{stoco}

\begin{figure}[htb]
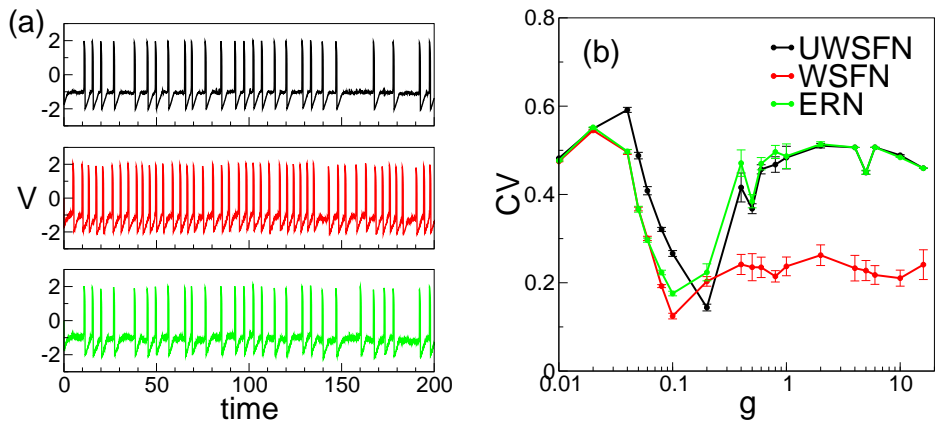

\raggedleft
\includegraphics[height=0.35\textwidth,keepaspectratio,clip]{figs/spikes}~~~
\includegraphics[height=0.35\textwidth,keepaspectratio,clip]{figs/CV_vs_g}
\caption{(a) Spiking dynamics of the unweighted SFN (top), weighted SFN
(middle) and random network (bottom) for noise intensity $D=0.5$ and
coupling strength $g=1$. (b) Coefficient of variation (CV) versus coupling strength
for same noise and the three network types mentioned above.
}
\label{fig:2}
\end{figure}

We now turn to the spiking activity of the networks discussed above, and ask
whether the increased synchronizability exhibited by the weighted SFNs concurs
with a decreased response to noise of the collective dynamics of the system.
Figure~\ref{fig:2}(a) shows sample time traces of the three networks in the pulsing
regime for a fixed noise intensity and coupling strength. In order to 
quantify in a systematic way the regularity of this spiking dynamics, we analyze
the distribution of time intervals between pulses, and in particular we calculate
the normalized standard deviation (also known as coefficient of variation)
of that distribution, $CV = \left\langle\sigma/T\right\rangle$,
where $T$ and $\sigma$ are the temporal average and standard deviation
of those intervals, respectively, and the brackets denote
the average over all nodes in the network and over the whole set of network replicas
(ten for each network type).

The dependence of CV on the coupling strength is seen in Fig.~\ref{fig:2}(b).
All three network classes show a clear minimum of the coefficient of variation
for an intermediate level of coupling strength, which is a signature
of array-enhanced coherence resonance \cite{changsong,nuestro}:
an optimal amount of coupling improves the coherent behavior of 
the system. Notably, the regularity is larger (CV is smaller) for the
{\em weighted} SFN than for the other two networks for almost all
coupling strengths (with the exception of the optimum coupling for the
unweighted SFN, which incidentally occurs at a larger value than
the weighted case). The difference is specially evident for larger
coupling strengths, where CV is less than half for the weighted SFNs
than for the other two networks. Thus, not only weighted SFNs
synchronize better than the other two complex network architectures,
but they also respond better to noise.

\begin{figure}[htb]
\centering
\includegraphics[height=0.35\textwidth,keepaspectratio,clip]{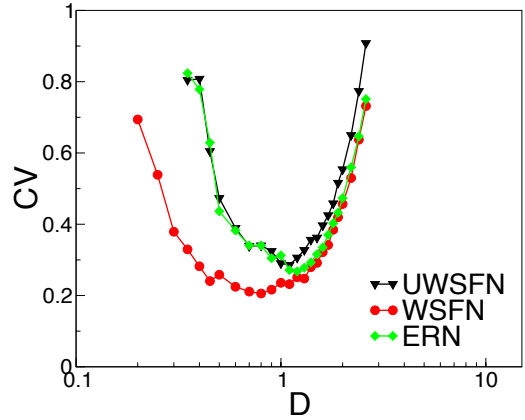}
\caption{Coeficient of Variation (CV) as a function of noise amplitude (D)
showing stochastic coherence for a fixed coupling strength ($g=1$)
and for the three networks discussed in the text.
}
\label{fig:3}
\end{figure}

In array-enhanced coherence resonance, regular behavior is also
enhanced for an optimal noise intensity. This stochastic coherence
effect is shown in Fig.~\ref{fig:3}, which represents the coefficient
of variation versus noise intensity for the three different types of network
discussed above. 
Similarly to the behavior shown with respect
to the coupling strength, CV is here lower for the weighted SFN than
for both the unweighted SFN and random network (which are very
much alike to one another), for all noise intensities, the difference being
most noticeable for low noise. Furthermore, the optimal noise level is
smaller in the weighted SFNs. In consequence, we can conclude
that weighted SFNs show both a better synchronizability and a superior
collective response to noise.

\begin{figure}[htb]
\raggedleft
\includegraphics[height=0.35\textwidth,keepaspectratio,clip]{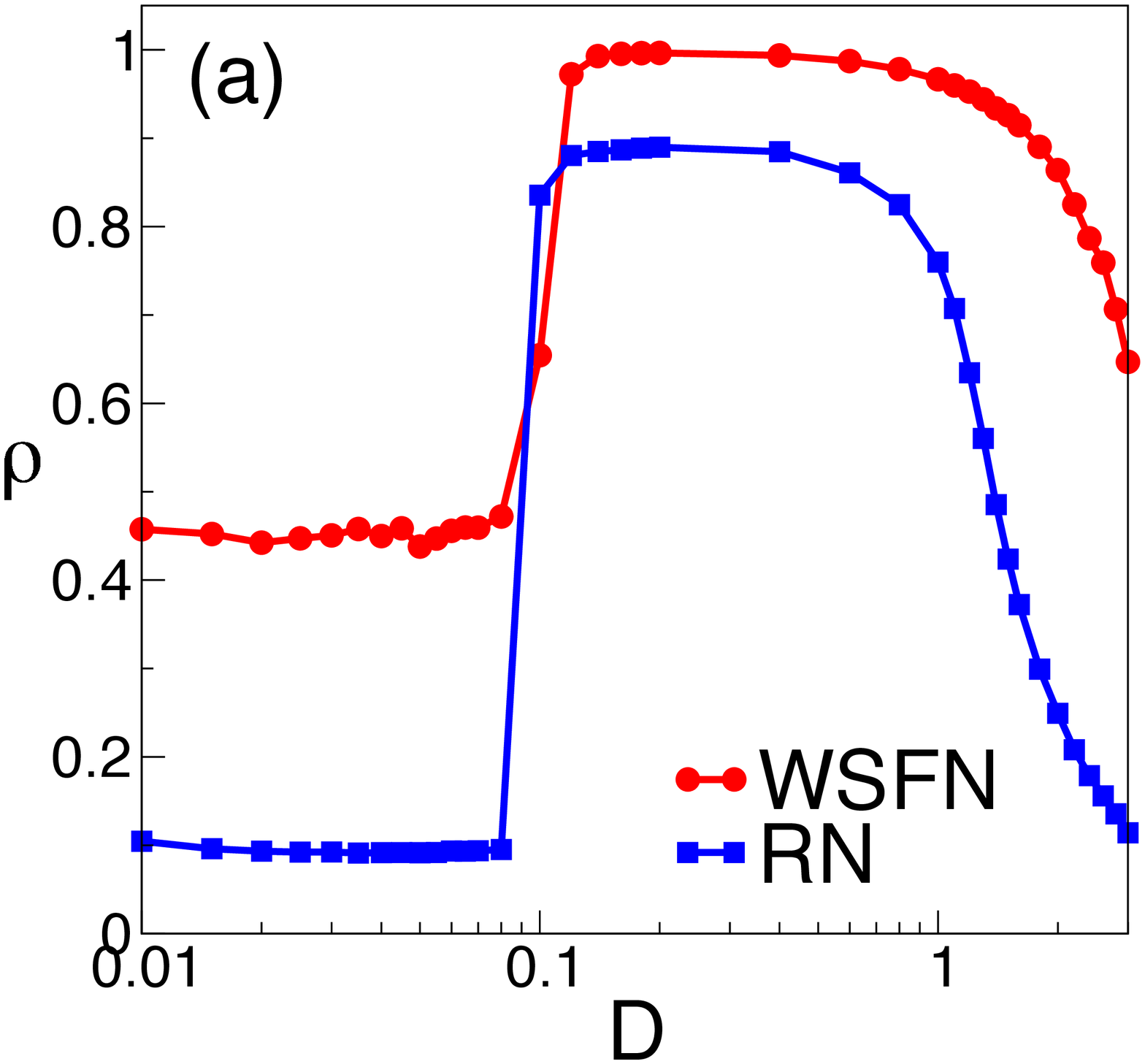}~~~
\includegraphics[height=0.35\textwidth,keepaspectratio,clip]{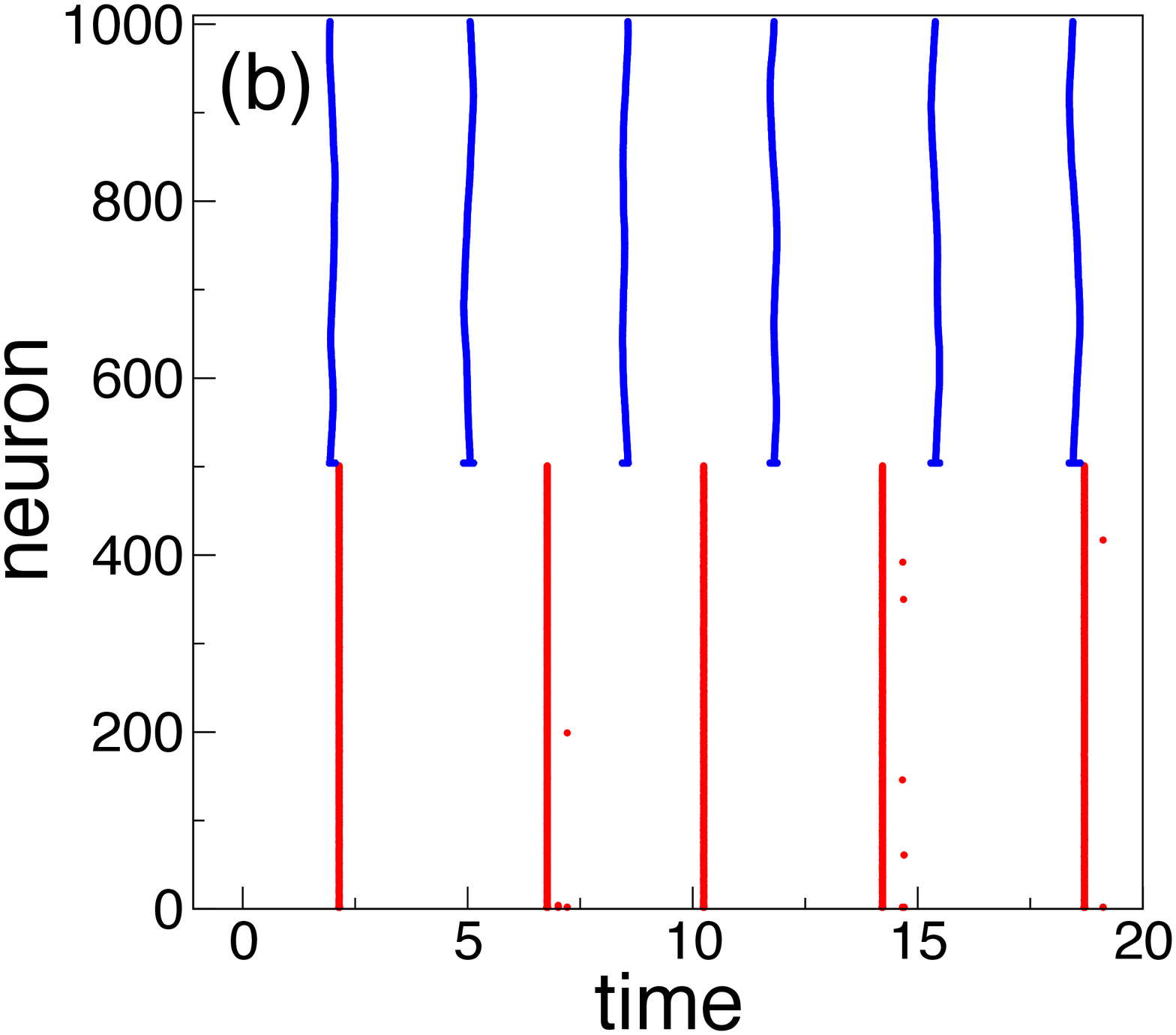}
\caption{(a) Synchronization level, measured via the synchronization coefficient $\rho$,
versus noise intensity $D$, for a weighted SFN (red, circles) and a regular network
(blue, squares), and for $g=1$. (b) Raster plots showing the location of the pulses in
$x_i$ for the same two networks with $g=1$ and $D=0.5$.
}
\label{fig:4}
\end{figure}

We have not discussed so far how the behavior of the weighted SFN
compares with that of a regular network (i.e. a network with only nearest-neighbor
coupling between its elements). Due to the lack of long-range coupling,
regular networks synchronize very poorly, as shown in Fig.~\ref{fig:4}.
Panel (a) in that figure plots the synchronization coefficient defined
in Sec.~\ref{sync} above with respect
to noise intensity, for a wide range of noise levels covering both
the subthreshold and spiking regimes. The transition between the two
regimes can be identified by the sudden increase in $\rho$ occurring
at $D\sim 0.1$.
The figure shows that for almost all noise levels, corresponding to both the subthreshold
and spiking regimes, the synchronization is substantially larger for the weighted SFN
than for the regular network. The raster plot in Fig.~\ref{fig:4}(b) reveals that the low
synchronization of the regular network is due to the finite propagation time of the
excitations throughout the network, in comparison with the basically instantaneous
propagation enabled by the long-range connections in the weighted SFN. Therefore,
the latter type of network is also superior to regular networks in optimizing {\em both}
synchronization and collective noise response simultaneously.

\section{Correlating synchronizability and stochastic coherence with degree}

\begin{figure}[htb]
\raggedleft
\includegraphics[height=0.35\textwidth,keepaspectratio,clip]{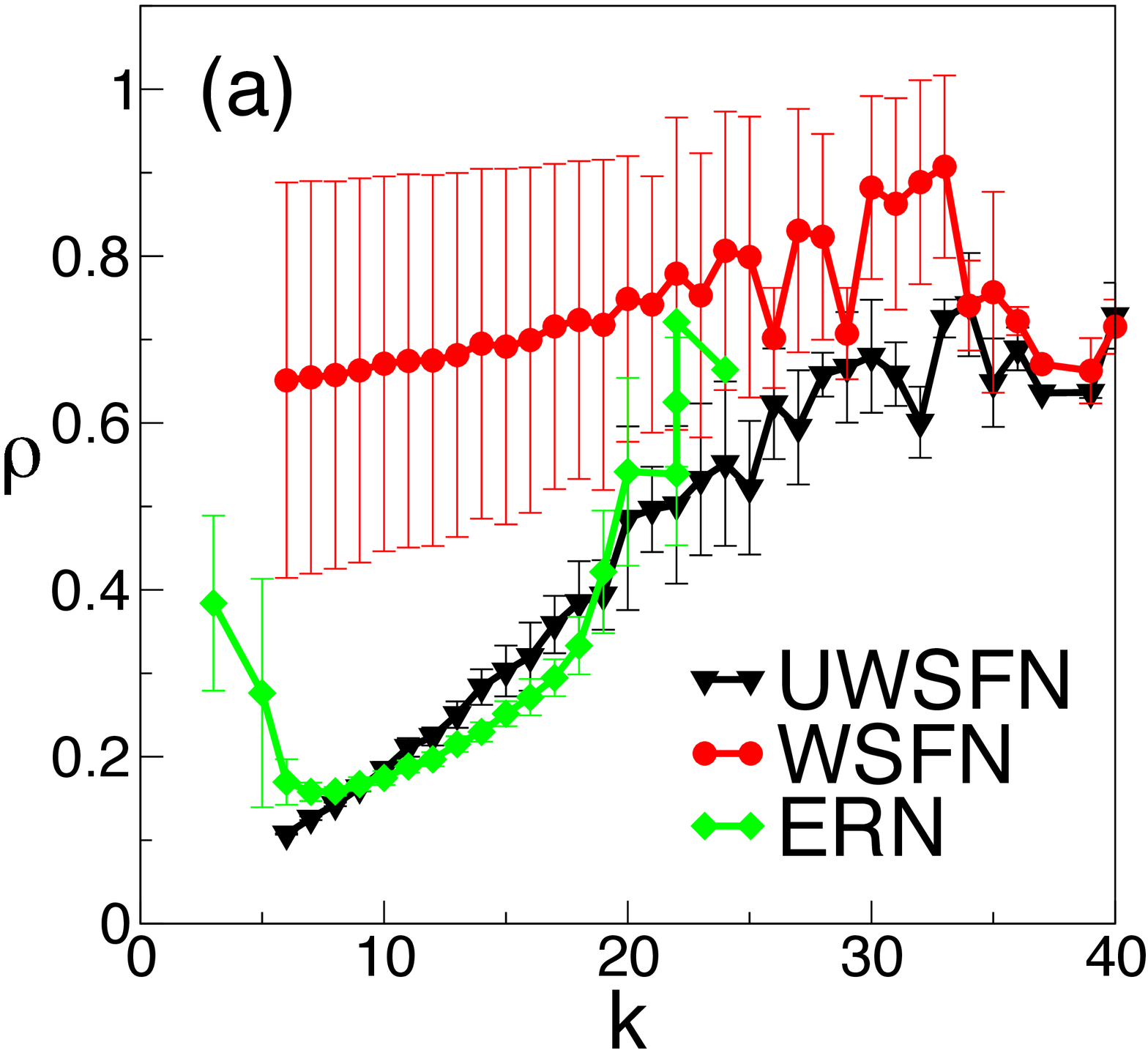}~~~
\includegraphics[height=0.35\textwidth,keepaspectratio,clip]{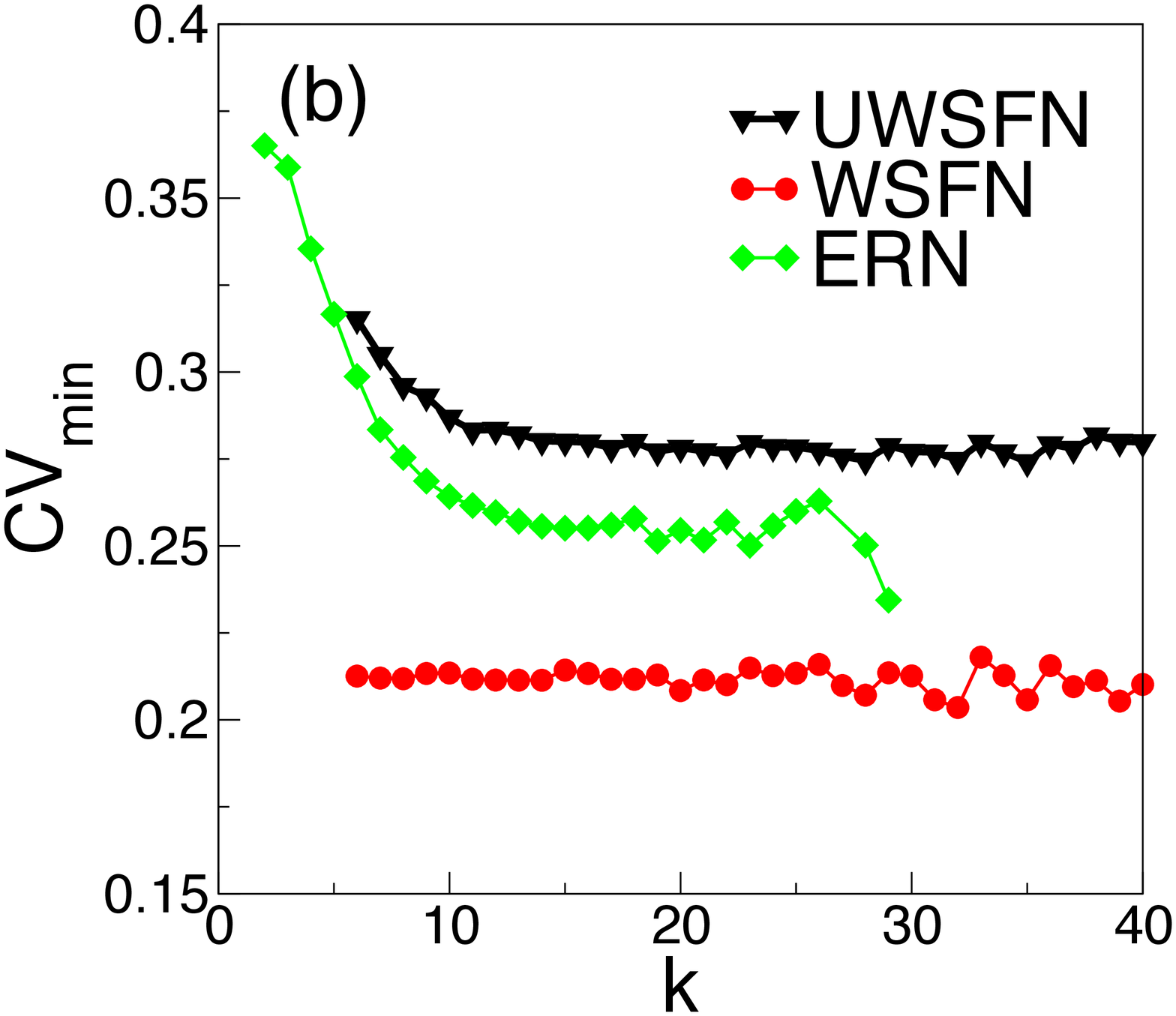}
\caption{Synchronization level at $D=0.1$, measured via the synchronization coefficient
$\rho$ (a), and minimum value of the coefficient of variation $CV_{min}$ (b), both as a
function of the degree node $k$ for the unweighted SFN (black triangles), weighted SFN
(red circles) and random network (green diamonds) with $g=1$.}
\label{fig:5}
\end{figure}

In order to investigate the mechanism behind the dual optimality of weighted SFNs
with respect to both synchronizability and collective stochastic coherence,
we now examine in detail how these two properties vary in nodes with different
degree. First we plot in Fig.~\ref{fig:5}(a) the synchronization coefficient
$\rho$ for varying node degree $k$, again comparing the weighted SFN, unweighted
SFN, and random network. The figure shows that synchronization 
increases basically monotonically
with the degree in all three cases, since higher connected nodes will be more strongly
synchronized. However, the dependence of $\rho$ on $k$ is much weaker in the
case of the weighted SFN, which reflects the compensating effect that the
coupling strength normalization given by Eq.~(\ref{newCouplingScheme})
has on the coordination between pairs of nodes: when two such coupled
nodes have small degrees (which would diminish their synchronization),
their connection becomes more important for
the global topological structure of the network, thus increasing the coupling
strengths between them (and so enhancing synchronization between them). 
As a consequence, the {\em average} synchronization level becomes larger
for this type of networks than for unweighted and random networks (Fig.~\ref{fig:1}).

The dependence of the regularity on the node degree is even more revealing.
Figure~\ref{fig:5}(b) shows the minimum (with respect to noise) of the local
coefficient of variation for different node degrees. `Local' here
refers to the fact that the CV is computed only for nodes with a given $k$. This
figure shows a clear decrease of CV$_{\rm min}$ for
both unweighted SFNs and random networks: nodes with low connectivity are
substantially less regular than nodes with high connectivity in these networks.
In contrast, this behavior is completely absent in weighted SFNs, where
CV$_{\rm min}$ is basically independent of the degree (and much smaller overall,
as noted also in Sec.~\ref{stoco} above). Once again, the coupling strength
normalization provided by the weighting of the links balances the disordering
effects of having a low degree, thus compensating perfectly the effects of
topology heterogeneity, and leading to a homogeneous coherence throughout
the network, which results in an enhanced averaged coherence.

\section{Discussion and conclusions}

Synchronization and collective noise response are in principle opposing
phenomena, since array-enhanced coherence resonance requires a certain
amount of dynamical heterogeneity: in the limit of perfect synchronization
the system behaves as a single unit and coupling would have no
effect on noise-induced coherence. Thus it should be expected that
systems that synchronize well (such as standard scale-free networks
or random networks) have poor collective stochastic coherence,
whereas systems that do not synchronize perfectly (such as regular networks,
in which activity waves propagate spatially with finite speed) can
respond positively to noise in terms of their regularity \cite{changsong}.
The results above show that certain weighted scale-free networks
exhibit both high synchronizability and a large level of
stochastic coherence induced by coupling. The weighting process
to which the links are subjected in those networks reduces the
heterogeneity to a level for which synchronization is now
possible, while array-enhanced stochastic coherence is not lost. For that reason,
we conjecture that the weighted scale-free networks presented here
have an optimal coupling topology for collective operation in
stochastic environments.

\section*{Acknowledgments}

This work was supported by the Ministerio de Economia y Competividad
(Spain, project FIS2012-37655),  the Generalitat de Catalunya (project 2009SGR1168),
 CONICET (PIP:0802/10), and University of Buenos Aires (UBACyT 20020110200314).
JGO acknowledges support from the ICREA Academia programme.

\section*{References}

\bibliographystyle{ieeetr}

\bibliography{crsfn}

\end{document}